# A Proposal for Energy-Efficient Cellular Neural Network based on Spintronic Devices

Chenyun Pan and Azad Naeemi
School of Electrical and Computer Engineering,
Georgia Institute of Technology, Atlanta GA 30332

**Abstract**
Due to the massive parallel computing capability and outstanding image and signal processing performance, cellular neural network (CNN) is one promising type of non-Boolean computing system that can outperform the traditional digital logic computation and mitigate the physical scaling limit of the conventional CMOS technology. The CNN was originally implemented by VLSI analog technologies with operational amplifiers and operational transconductance amplifiers as neurons and synapses, respectively, which are power and area consuming. In this paper, we propose a hybrid structure to implement the CNN with magnetic components and CMOS peripherals with a complete driving and sensing circuitry. In addition, we propose a digitally programmable magnetic synapse that can achieve both positive and negative values of the templates. After rigorous performance analyses and comparisons, optimal energy is achieved based on various design parameters, including the driving voltage and the CMOS driving size. At a comparable footprint area and operation speed, a spintronic CNN is projected to achieve more than one order of magnitude energy reduction per operation compared to its CMOS counterpart.

## 1. Introduction

CMOS scaling has lasted for half a century and almost approaches its fundamental and physical limits [1]. Many novel device technologies have been proposed in the past decade to replace or augment the conventional CMOS technology [2]. To further improve the computing energy efficiency, spintronic devices are developed using electron spin as a state variable to perform logic computation [3-5]. However, despite the large research efforts in the Boolean logic domain, few device concepts have better or comparable performance or energy efficiency than the conventional CMOS technology [2]. Some non-Boolean computing architectures, such as the cellular neural network (CNN) [6], are potential candidates to provide higher energy efficiencies due to their massive parallel processing capability. The CNN also has the advantage of its compatibility with integrated circuits and has a wide range of applications in image and signal processing, such as the pattern recognition and motion tracking [7-9].

A CNN contains an array of computing cells that are connected only with nearby cells. Since interconnects are major limitations in modern VLSI systems, CNN systems can take the advantage of the local communication and encounter small constraints imposed by interconnects. The CNN can be considered as a brain-inspired computing architecture that is based on neurons to integrate the incoming currents. The accumulated and activated output signal drives nearby neurons through weighted synapses. The underlying mathematics of a CNN was proposed by Chua [6] and the dynamic state equation of each CNN cell circuit is written as

$$C\frac{dx_{ij}}{dt} = -\frac{1}{R}x_{ij} + \sum_{kl\in S_{ij}} A_{ij,kl}y_{kl} + \sum_{kl\in S_{ij}} B_{ij,kl}u_{kl} + I_{ij}, \qquad y_{ij} = f(x_{ij}) \qquad (1)$$

where $x_{ij}$ is the state voltage of the cell, $R$ and $C$ are linear resistance and capacitance of each cell, $y_{kl}$ and $u_{kl}$ are the outputs and inputs of neighboring cells, respectively, $f(x)$ is the sigmoid function that describes the characteristic between the output voltage and cell state voltage, $A_{kl}$ and $B_{kl}$ are templates of each cell, whose values represent the weights of synapses connecting two nearby cells, and $I_{ij}$ is the input bias current of each cell. Due to its massive symmetric and identical structure with local connections, the CNN can be easily implemented with analog VLSI circuits using operational amplifiers and operational transconductance amplifiers (OTAs) as neurons and synapses [10-12]. The main drawbacks of this analog implementation are large area and power dissipation.

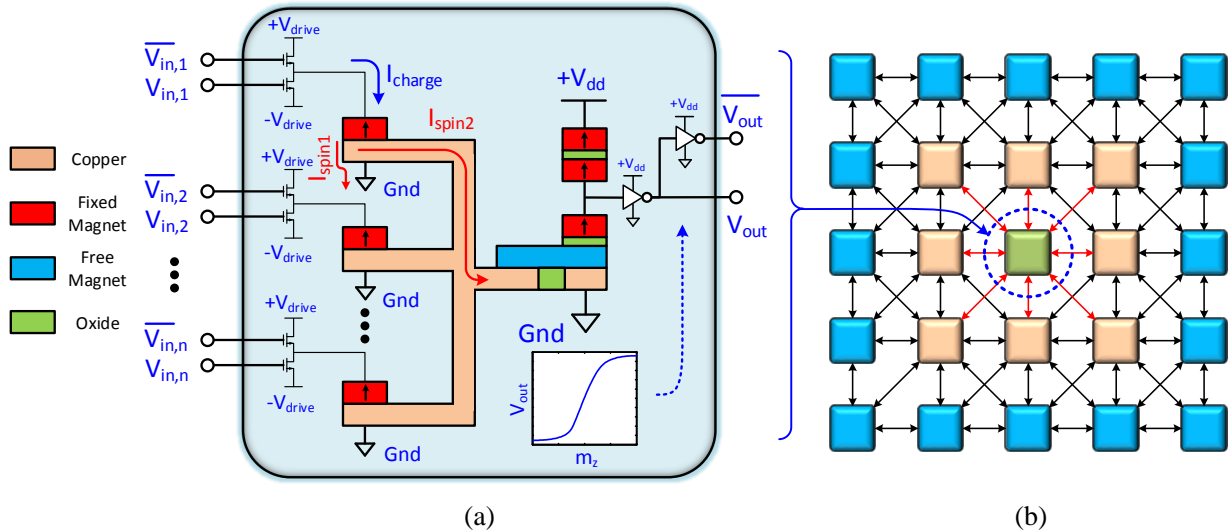

(a)                  (b)

Fig. 1. Illustrations of a CNN implemented with magnetic synapses and neurons. Figure (a) shows the cell diagram that includes magnetic synapses and a neuron. Figure (b) shows the connectivity of CNN, where each cell is only connected with its nearest neighbors.

In this paper, we propose an alternative way to implement the CNN using single-domain nanomagnets to reduce the area and improve the energy efficiency. The basis of this approach is that a magnet can be switched by the spin transfer torque induced by a spin-polarized current [13]. The amplitude of the spin-polarized current determines the switching delay of the magnet, which essentially makes the magnet act as an integrator. The basic building block is adopted from the all-spin logic (ASL) [5], which was proposed as one of the beyond-CMOS device options for the Boolean logic computation. Recent results from a benchmarking study for a 32-bit adder implemented by various emerging computational devices have shown that ASL is orders of magnitude slower and more power-hungry compared to CMOS [2]. Here, however, we demonstrate that implementing CNN with ASL devices is much more efficient, and the improvement is shown to be application dependent.

The rest of the paper is organized as follows. Section II describes the design methodology of a spintronic CNN, including the circuit and system design scheme and the modeling and simulation approach. Section III shows the simulation results for two widely used application. Performance benchmarking is performed for both spintronic and the conventional CMOS implementation in terms of delay, energy, and footprint area. Conclusions are made in Section IV.

## 2. Spintronic CNN Design Methodology

### A. *Circuit and System Design*

Fig. 1 illustrates a CNN implemented with magnetic synapses and neurons. The inputs of the CNN cell, $V_{in}$ and $\overline{V_{in}}$, come from both inputs, $u_{kl}$, and outputs, $y_{kl}$, of nearby cells, which are connected to n-type driving MOSFETs with various driving sizes determined by the absolute values of templates $A_{ij}$ and $B_{ij}$. Depending on the input voltage, one of the driving transistors is turned on, generating a charge current. This charge current gets spin-polarized, and the injected (or extracted) spins diffuse towards the free magnet (the neuron). The polarity of the spin current depends on the magnetization direction of the magnet and the charge current direction. For instance, if the input, $V_{in}$, is logic zero, the charge current flows from $+V_{drive}$ through the input magnet into the ground connection underneath the magnet, as shown by the blue arrow in Fig. 1 (a). Electrons with spins that are opposite to the magnetization direction of the magnet accumulate underneath the magnet. This spin accumulation creates a spin-polarized current, which separates into two paths, $I_{spin1}$ and $I_{spin2}$, toward the ground and the output free magnet, respectively, as shown by red arrows in Fig. 1 (a). These two spin-polarized currents satisfy the spin drift-diffusion equation [14]. For CNN cells with multiple inputs, spin-polarized currents from various input magnets superimpose, achieving the integration functionality. The overall net current inserts a spin torque into the output magnet, setting the magnetization direction of the free magnet in the neuron. Note that the amplitude of voltage source, $V_{drive}$, can be smaller than the $V_{dd}$ to allow low-power operation and improve the energy efficiency. The magnet magnetizations are preset to implement signs of template values.

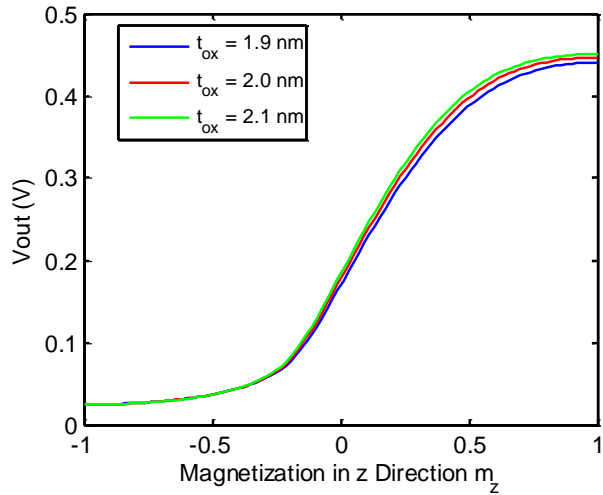

Fig. 2. Output voltage of the inverter versus the magnetization of the magnet in z direction at various oxide thickness of the MTJs.

The magnetization direction of the output magnet is read by building two magnetic tunneling junctions (MTJs). The top MTJ is set as the reference MTJ with a parallel configuration. For the bottom MTJ, if the magnetization of the free magnet is in the same direction as that of the fixed magnet, it has a parallel configuration and a small resistance, lowering the voltage between two MTJs; if the magnetization of the free magnet is in the opposite direction, a high voltage is generated. Since the output voltage of each cell needs to switch driving transistors in multiple synapses in nearby cells, an inverter is added at the output to amplify the voltage between two MTJs and provide a large voltage swing between the ground and $V_{dd}$. This sensing scheme does not need a sense amplifier, which is different from the STTRAM. The reasons are that 1) there are no extra resistance associated with selectors and long interconnects in the read path and 2) the oxide thickness of the MTJs can be larger than that of an STTRAM due to the separation between read and write paths. For an MTJ with a larger oxide thickness, not only the tunneling magnetoresistance (TMR) increase but also the absolute values of the parallel and antiparallel resistances increase exponentially. The read MTJ and the reference MTJ are matched with similar oxide thicknesses and therefore the output voltage is not sensitive to small changes in the oxide thickness which may happen due to the process variation. This is demonstrated in Fig. 2 where the output voltage of the inverter is plotted versus the magnet magnetization in z direction (out-of-plane) for three MgO thicknesses of 1.9, 2.0 and 2.1 nm. The corresponding MTJ parameters are from the experimental data reported in [15]. Since the sensing current from $V_{dd}$ to ground is much smaller compared to the critical switching current, the output magnet will not be disturbed during the read.

## B. *Modeling and Simulation Approach*

Magnet switching dynamics follow the Landau-Lifshitz-Gilbert (LLG) equation with a spin-transfer-torque term [13, 16]. The magnetization of a magnet, $\vec{m}$, under a perpendicular spin-polarized current, $\vec{I}_{S,\perp}$, is

$$\frac{d\vec{m}}{dt} = -\gamma\mu_0[\vec{m} \times \vec{H}_{eff}] + \alpha\left[\vec{m} \times \frac{d\vec{m}}{dt}\right] + \frac{\vec{I}_{S,\perp}}{qN_s} \quad (2)$$

where $\vec{H}_{eff}$ is the effective field, $\gamma$ is the gyro ratio, $\mu_0$ is the permittivity, $\alpha$ is the damping factor, $q$ is the elementary charge, $N_s$ is the number of magnetons, and $\vec{I}_S$ is the spin-polarized current. $\vec{I}_{S,\perp}$ can be expressed as $i_0 \cdot \left(\sum_{kl \in S_{ij}} A_{ij,kl} y_{kl} + \sum_{kl \in S_{ij}} B_{ij,kl} u_{kl} + I_{ij}\right)$, where $i_0$ is the unit spin-polarized current when the template value is unity. The amplitude and the direction of the spin-polarized current depend on the output and input voltage polarities of nearby cells and the weights of synapses connecting those cells.

Magnets used in the neurons are assumed to have perpendicular magnetic anisotropy with the dimensions of 30 nm × 30 nm × 2 nm, the saturation magnetization of 5×10$^5$ A/m, the perpendicular magnetic anisotropy of 6×10$^4$ J/m$^3$, and the damping coefficients of 0.01. The spin accumulation, $\mu_s$, satisfies the drift-diffusion equation $\partial^2 \mu_s / \partial x^2 = \mu_s / l_{sf}$, and the spin-polarized current $J_s = \sigma/e \cdot \partial \mu_s / \partial x$, where $l_{sf}$ is the spin relaxation length of copper calculated as 420 nm, which is longer than the distance between the input and output magnets of 100 nm, $\sigma$ is the interconnect conductivity, and $e$ is the elementary charge. Boundary conditions are as follows: the spin accumulation underneath the output magnet is zero, and the input spin-polarized current underneath the input magnet is $\beta \cdot I_c$, where $I_c$ is the charge current and the spin injection coefficient $\beta$ is 0.5. This value is extracted to match the delay obtained from the numerical simulation of complete spin injection and diffusion transportation mechanisms based on the modified nodal analysis with spin conductance matrix, which takes into account the contribution of ferromagnet/Cu interface [17]. For a CNN cell with multiple inputs, the spin-polarized current $I_{spin2}$ from one input may diffuse to the rest through parasitic paths. However, since the boundary condition of the spin accumulation at the output magnet is zero, majority of the spins are absorbed at the output magnets instead of the other input magnets. The ground path resistance is set as 50 Ω. The thermal noise is considered at the room temperature of 300 K. The MTJs used at the output of each neuron have parallel and antiparallel resistances that are taken from the data reported in the experiment [15] with a tunneling oxide, MgO, thickness of 2 nm, which corresponds to a TMR of 160%. For the CMOS circuitry, the IV characteristic of inverters and driving transistors are simulated in HSPICE with the Arizona State University Predictive Technology Model at the 16 nm technology node [18].

## 3. Simulation Results

### A. *Functionality Demonstration*

Based on Fig. 1, two CNN applications are demonstrated: noise filtering and associative memory. The first one uses a fixed weight implementation, and the latter one has synapses that are programmable in accordance to the training pattern sets.

#### *1) Noise Filtering*

Noise filtering is one of the most common applications in image processing. To demonstrate the functionality, a pattern '0', consisting 30×20 pixels with 10% random noise, is given as the initial state of the CNN. As shown in Fig. 3 (b), black and white pixels represent the magnetization directions of magnets pointing up and down, respectively. The templates for the noise filtering application implemented with spintronic devices are set as

$$A = \begin{bmatrix} 0 & 1 & 0 \\ 1 & 1 & 1 \\ 0 & 1 & 0 \end{bmatrix}, B = 0, I = 0 \quad (3)$$

Since there is no intrinsic self-feedback in the spintronic CNN, the middle element in template A is "1" rather than "2" compared to the CMOS implementation, where the circuit has a linear feedback resistance $R$ in the dynamic equation (1). To correctly filter the noisy pixels, the net spin-polarized current received in the magnets representing noisy pixels needs to be larger than the critical switching current. For the demonstration shown in Fig.

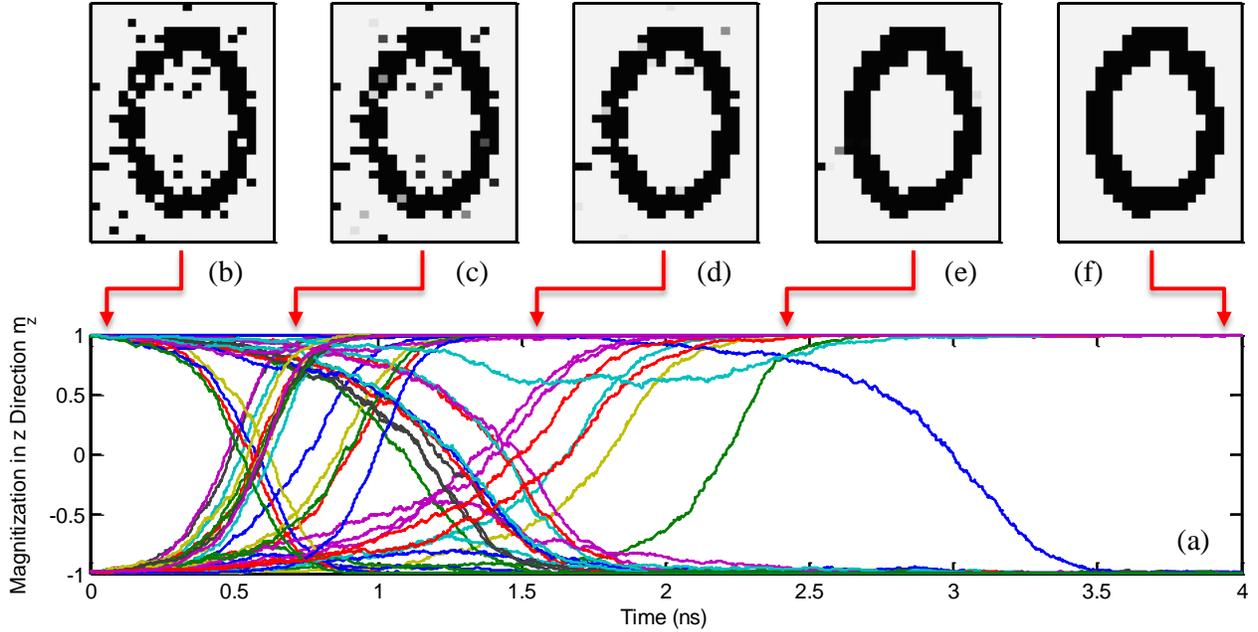

Fig. 3. Functional demonstration of a spintronic CNN performing the noise filtering application. Figure (a) shows the transient response of the magnet magnetization in z direction, where curves with different colors represent the magnetization in various cells. Figures (b) - (f) show the corresponding output patterns at t = 0, 0.7, 1.5, 2.4, and 4 ns, respectively.

3, the spin-polarized current corresponding to the unit weight is set as ten times of the critical switching current. In the next subsection, this current is varied to minimize the overall energy dissipation by optimizing the driving voltage and driving peripherals. As the simulation starts, magnets corresponding to noisy pixels receive spin torques whose directions are opposite to magnet magnetizations, and these magnets switch to the opposite directions. Fig. 3 (a) shows the transient response of the magnetization of switching magnets in z direction. Fig. 3 (b) – (f) show five snapshots of the pattern evolving with time. The final output after 4 ns is displayed in Fig. 3 (f), where noisy pixels are filtered correctly. This simulation was performed at the room temperature of 300K, and the thermal noise in the magnets was included in simulations.

*2) Associative Memory*

Another popular application that can be effectively implemented with a CNN is the associative memory, which has been widely used in the pattern and sound recognition [19, 20]. For this type of applications, the weights of synapses vary among cells depending on the patterns to be trained and memorized. Fig. 4 demonstrates an application with four training patterns. Patterns '1' and '3' are associated with patterns '2' and '4', respectively, as shown in Fig. 4 (a). The training method used here follows Hebbian learning algorithm [7]. With a fair computational cost and a good convergence speed, it can be applied for a large number of free parameters in a space-varying template used in the associative memory. After the training is complete, templates *A*, *B*, and input bias currents *I* are obtained for each cell. To test the functionality, a pattern '1' with four noisy pixels is given as both the initial states and the inputs of magnets in the cell neurons. Once the simulation starts, magnets in neurons receive spin-polarized currents from nearby cell outputs, inputs and self-bias currents. Fig. 4 (b) shows the transient response of the magnetization of switching magnets in the z direction. Fig. 4 (c) – (h) show five snapshots of the pattern evolving over time. The final output after 10 ns is displayed in Fig. 4 (h), where a correct associated pattern '2' is realized.

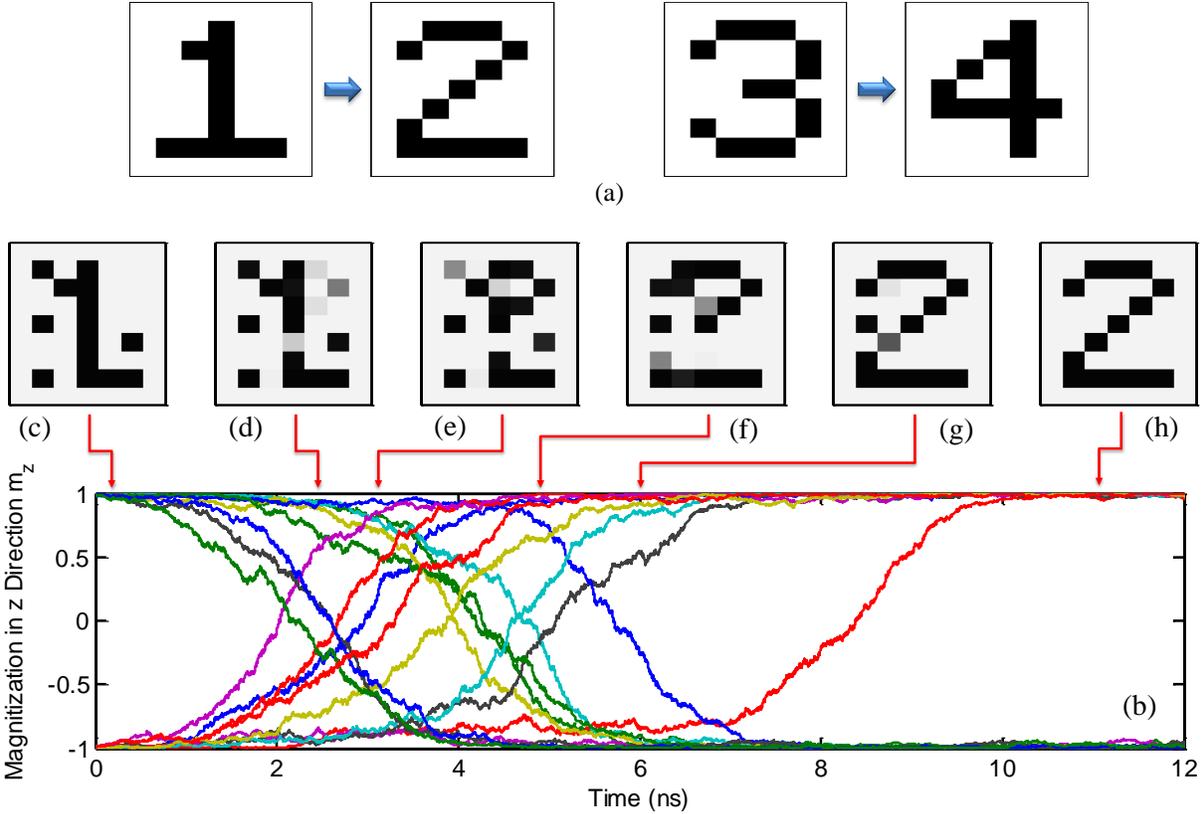

Fig. 4. Functional demonstration of a spintronic CNN performing the associative memory application. Figure (a) shows the training sets, where patterns '1' and '3' are associated with patterns '3' and '4'. Figure (b) shows the transient response of the magnet magnetization in z direction, where curves with different colors represent the magnetization for various cells. Figures (c) – (h) show the corresponding output patterns at t = 0, 2.2, 3, 4.5, 6, and 11 ns, respectively.

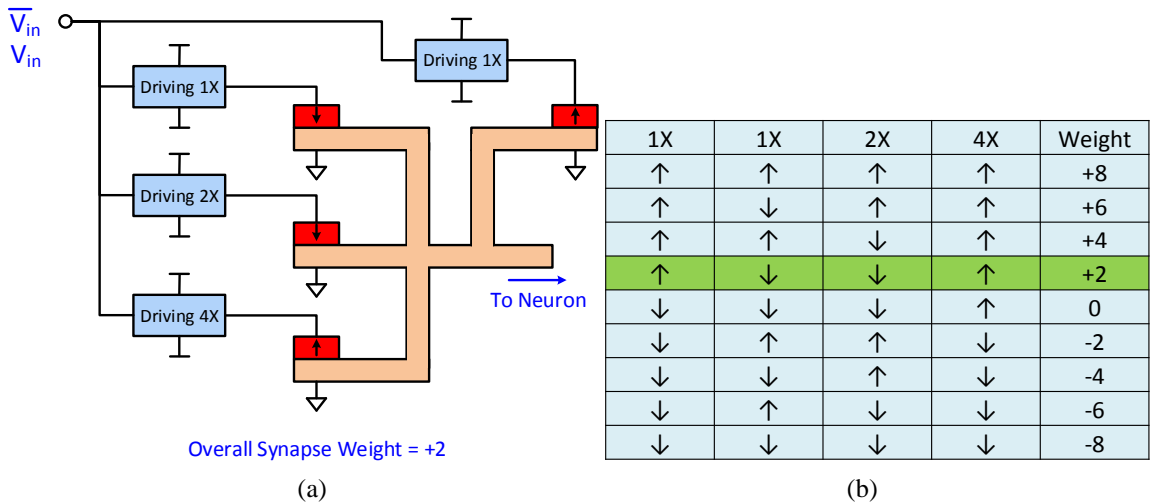

Fig. 5. A magnetic synapse that has a digitally programmable weight. The driving transistor size is quantized as shown in (a). (b) shows various combinations of input magnets and corresponding weights, where the green line is illustrated in (a).

For cases in which the weights of synapses need to be programmable, multiple magnets with quantized driving sizes are implemented. Fig. 5 shows an example of a synapse whose weight can be digitally programmable. Four magnets are used for this synapse, and the corresponding driving transistor sizes are set as 4×, 2×, 1×, and 1× of the minimum size. Different input weights, including negative weights, are achieved by setting the magnetization direction of the four magnets according to Fig. 5 (b), where the fourth configuration (green line) is illustrated in Fig. 5 (a). For a higher resolution requirement, the number of magnets and driving transistors increases, which leads to a larger footprint area. To program the magnets, an extra layer of fixed magnet is placed on top of the input magnets. Depending on the voltage polarity applied on the fixed magnet, the input magnet is programmed to be in the parallel or antiparallel configuration with the fixed magnet.

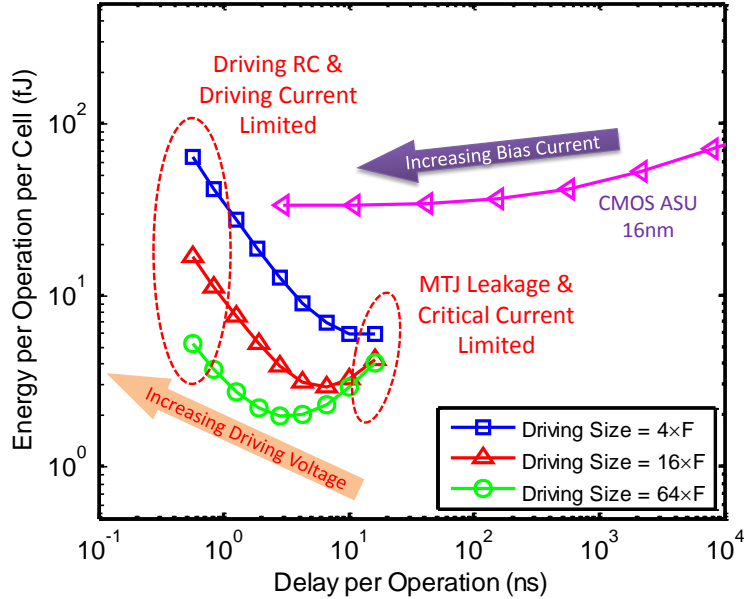

Fig. 6. Performance comparison between spintronic and CMOS CNNs in terms of energy per operation and delay, where four different driving sizes of spintronic CNN are investigated.

## B. *Performance Analyses and Comparison with the CMOS Technology*

For the performance analyses of a CNN implemented by the CMOS, neurons are built with two-stage differential-input single-ended output 7-transistor operational amplifiers [21], and synapses are realized with OTAs [12]. The power consumption of each component, including operational amplifiers and OTAs, are simulated based on SPICE simulations using the same CMOS model in the spintronic CNN at the 16 nm technology node [18]. Tail transistors in different OTAs have quantized widths, and the gates of the tail transistors are connected to memory cells to make the weight of the synapses digitally programmable. The dynamic state voltages of the CNN are simulated in MATLAB with equation (1), where the linear feedback capacitance $C$ is the summation of the input and output capacitances of nearby OTAs to reliably sink current. The power-delay scaling analyses of various weights of synapses and neurons are adopted from previous work [12], where the linear feedback resistance $R$ scales down as the bias currents of amplifiers increase so that the energy-delay trade-offs can be realized. The footprint area of operational amplifiers and OTAs are estimated by the summation of widths of all transistors multiplied by 8F, where F is the minimum feature size of 16 nm. This may underestimate the analog circuit area overhead, but considering the fact that the results are shown in log scale in Fig. 7, it provides a reasonable approximation in an efficient way.

### *1) Energy and Delay Optimization*

Fig. 6 compares the performances of a spintronic CNN and its CMOS counterpart. To explore the energy and delay trade-offs for the spintronic CNN, three different driving sizes are investigated. The read voltages for MTJs and inverters are set as 0.7V, and the driving voltage of synapses is varied between 10mV to 1V, providing the energy and delay trade-off as shown in Fig. 6. The corresponding charge current passing through the magnet varies between 2.8 μA and 75 μA.

For a specific noise filtering application with a given noise, shown in Fig. 3 (a), the delay of the spintronic CNN keeps decreasing as the driving voltage increases because the larger spin-polarized current density reduces the magnet switching time. For the energy dissipation, an optimal driving supply voltage exists to minimize the energy. This is because when the driving voltage is high, reducing the voltage lowers the charge current and the energy dissipation associated with the joule heating of driving transistors; however, if the driving voltage becomes too small, the energy dissipation starts to increase. This is mainly due to two reasons: 1) the spin-polarized current keeps decreasing and approaches the critical switching current of the magnet in the neuron, causing the switching time of the magnet to increase significantly; and 2) a constant leakage current path in the sensing MTJs and inverters contributes to the leakage energy. Since the energy is the product of power and delay, the power saving no longer compensates the rapid increase in the delay, leading to a jump in the overall energy dissipation. To further reduce the energy dissipation, we increase the driving transistor size and reduce the resistance that the current needs to pass through. Energy reduction is achieved at the cost of an enlarged footprint area.

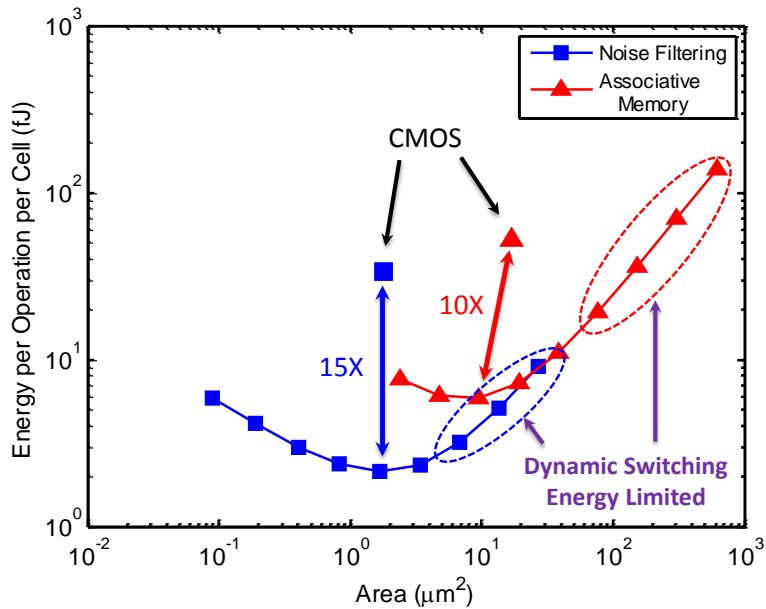

Fig. 7. Energy per operation versus footprint area for CMOS and spintronic CNNs. Blue and red curves show the trade-offs between energy per operation and footprint area for a spintronic CNN at minimum energy design points based on optimal driving voltages.

For a CNN implemented with the conventional CMOS technology, the bias currents of neurons and synapses are swept to study the energy and delay trade-off. The CNN dynamic is simulated in MATLAB based on equation (1), and the corresponding power and energy dissipation are obtained from SPICE simulations. As the bias current reduces, the delay increases along with the energy dissipation. The energy saturates as the delay approaches 10 ns, beyond which the CMOS amplifier is limited by the bandwidth and the DC gain of the amplifier, which affect the functionality of the CNN operation. Comparing CMOS and spintronic CNNs, the energy dissipation of the spintronic CNN is over one order of magnitude less at comparable operation speeds.

*2) Footprint Area and Energy Trade-offs*

Fig. 7 compares the footprint areas of CMOS and spintronic CNNs with minimum energy dissipation design points as shown in Fig. 6. As mentioned earlier, a larger driving size of the inverter reduces the energy dissipation at the cost of increasing the overall footprint area. Compared to the magnet size, the driving transistor area is much larger and dominates the overall footprint area of the spintronic CNN. However, if the driving size is too large, the switching energy associated with charging and discharging the large gate capacitance increases significantly, which eventually increases the overall energy per operation. Two applications are compared herein: noise filtering and associative memory applications. Compared to the noise filtering application, the associative memory application has a larger area consumption for both CMOS and spintronic CNNs because the synapse weight needs to be programmable based on the pattern sets and training processes. This requires multiple synapses with quantized weights, and the results shown here are based on a 3-bit resolution. The comparison between CMOS and spintronic CNNs shows that at comparable footprint areas, the spintronic CNN provides more than one order of magnitude improvement in the energy dissipation per operation. The reason for the smaller improvement observed in the associative memory application is as following. To achieve a small weight synapse, weighted spin-polarized currents in opposite directions cancel each other, and the remaining net current contributes to the switching current. This requires a larger driving current than the actual switching current, and subsequently increases the energy overhead.

### 3) *Implication and Discussion*

The design methodology proposed in this paper can be adopted by other spin-torque transfer mechanisms as well, such as the giant spin hall effect and the domain wall motion [22, 23]. Based on the present demonstration, although the individual magnet switching time is influenced by the thermal fluctuation, the magnet can take advantage of the noise tolerant property of a CNN. This reduces the switching current, brings down the driving voltage, and performs various tasks in an energy-efficient way.

One advantage of the spintronic CNN over the CMOS CNN is the area saving because small size magnets and inverters can replace the area consuming operational amplifiers and OTAs. The area of a spintronic CNN is dominated by inverters that are required to provide complimentary voltages and bi-directional output currents to drive magnetic synapses and generate the spin-polarized current. For a CNN designed with a small footprint area, the majority of energy dissipation is associated with the joule heating of driving currents passing through transistors; for a CNN designed with minimum energy per operation, the dynamic switching energy also has a non-negligible impact on the total energy dissipation, which depends on the switching activity factor, namely the number of pixels that are flipped during each operation.

Another advantage is that the synapse implemented with magnets can provide both positive and negative weights by simply reverse the initialization of the magnet magnetization orientations without adding extra resources or compromising the performance. For the CMOS synapse, an extra inverter, a multiplexer, and a memory unit are required to store the sign of the weight, which induce delay, energy, and area overheads. The other benefit of the magnet synapse is its non-volatility. The weights stored inside the magnet remain unchanged even during a power loss. This allows the spintronic system to be completely turned off during the idle state when no computation is running. However, for the CMOS CNN, the weights are commonly stored in volatile memory cells (e.g. SRAM) that control the tail transistors in OTAs to achieve the programmability. Therefore, the power supply for those memory cells needs to be kept on at all times. For a low-throughput application, the spintronic CNN has a considerable energy saving because the system can be turned off during the majority of the time.

## 4. Conclusions

In this work, an alternative way of implementing a CNN is proposed with magnets and complete CMOS peripheral sensing and driving circuits. The functionality is validated by performing a rigorous numerical simulation at a complete dynamic system scale. Performances of both the CMOS and the spintronic CNNs are investigated in two widely used applications: noise filtering and associative memory. Furthermore, a digitally programmable magnetic synapse is proposed to allow both positive and negative values of the templates. Comprehensive energy/delay/area trade-offs are investigated, and optimal design parameters are characterized to maximize the energy efficiency. A fair comparison is made based on the HSPICE simulation using the same technology models at the same node for both CMOS implementation and the peripheral circuits in the spintronic CNN. Compared with the CMOS technology, the spintronic CNN can potentially achieve more than one order of magnitude improvement in energy at a comparable footprint area, and this improvement is also demonstrated to be application dependent.

In addition, key limitations of the spintronic CNN are identified, including the footprint area and the dynamic switching energy of the peripheral circuits.

## Acknowledgement

This work was supported by the Semiconductor Research Corporation (SRC) NRI Theme 2624.001.